\documentclass[twoside,twocolumn,9pt]{article}

\usepackage{extsizes}
\usepackage[super,sort&compress,comma]{natbib} 
\usepackage[version=3]{mhchem}
\usepackage[left=1.5cm, right=1.5cm, top=1.785cm, bottom=2.0cm]{geometry}
\usepackage{balance}
\usepackage{mathptmx}
\usepackage{sectsty}
\usepackage{graphicx} 
\usepackage{lastpage}
\usepackage[format=plain,justification=justified,singlelinecheck=false,font={stretch=1.125,small,sf},labelfont=bf,labelsep=space]{caption}
\usepackage{float}
\usepackage{fancyhdr}
\usepackage{fnpos}
\usepackage[english]{babel}
\addto{\captionsenglish}{}
\usepackage{array}
\usepackage{droidsans}
\usepackage{charter}
\usepackage[T1]{fontenc}
\usepackage[usenames,dvipsnames]{xcolor}
\usepackage{setspace}
\usepackage[compact]{titlesec}
\usepackage[colorlinks=true]{hyperref}
\usepackage{amsfonts,amsmath,amssymb}
\usepackage{siunitx}
\usepackage{textcomp}
\newcommand{\ftextnumero}{{\fontfamily{txr}\selectfont \textnumero}}
\definecolor{cream}{RGB}{222,217,201}

\DeclareSIUnit\molar{\textsc{m}}
\usepackage{microtype}

\begin{document}

\pagestyle{fancy}
\thispagestyle{plain}
\fancypagestyle{plain}{
\renewcommand{\headrulewidth}{0pt}
}

\makeFNbottom
\makeatletter
\renewcommand\LARGE{\@setfontsize\LARGE{15pt}{17}}
\renewcommand\Large{\@setfontsize\Large{12pt}{14}}
\renewcommand\large{\@setfontsize\large{10pt}{12}}
\renewcommand\footnotesize{\@setfontsize\footnotesize{7pt}{10}}
\makeatother

\renewcommand{\thefootnote}{\fnsymbol{footnote}}
\renewcommand\footnoterule{\vspace*{1pt}%
\color{cream}\hrule width 3.5in height 0.4pt \color{black}\vspace*{5pt}} 
\setcounter{secnumdepth}{5}

\makeatletter 
\renewcommand\@biblabel[1]{#1}            
\renewcommand\@makefntext[1]%
{\noindent\makebox[0pt][r]{\@thefnmark\,}#1}
\makeatother 
\renewcommand{\figurename}{\small{Fig.}~}
\sectionfont{\sffamily\Large}
\subsectionfont{\normalsize}
\subsubsectionfont{\bf}
\setstretch{1.125} 
\setlength{\skip\footins}{0.8cm}
\setlength{\footnotesep}{0.25cm}
\setlength{\jot}{10pt}
\titlespacing*{\section}{0pt}{4pt}{4pt}
\titlespacing*{\subsection}{0pt}{15pt}{1pt}

\fancyfoot{}
\fancyfoot[LO,RE]{\vspace{-7.1pt}}
\fancyfoot[CO]{\vspace{-7.1pt}}
\fancyfoot[CE]{\vspace{-7.2pt}}
\fancyfoot[RO]{\footnotesize{\sffamily{1--\pageref{LastPage} ~\textbar  \hspace{2pt}\thepage}}}
\fancyfoot[LE]{\footnotesize{\sffamily{\thepage~\textbar 1--\pageref{LastPage}}}}
\fancyhead{}
\renewcommand{\headrulewidth}{0pt} 
\renewcommand{\footrulewidth}{0pt}
\setlength{\arrayrulewidth}{1pt}
\setlength{\columnsep}{6.5mm}
\setlength\bibsep{1pt}

\makeatletter 
\newlength{\figrulesep} 
\setlength{\figrulesep}{0.5\textfloatsep} 

\newcommand{\topfigrule}{\vspace*{-1pt}%
\noindent{\color{cream}\rule[-\figrulesep]{\columnwidth}{1.5pt}} }

\newcommand{\botfigrule}{\vspace*{-2pt}%
\noindent{\color{cream}\rule[\figrulesep]{\columnwidth}{1.5pt}} }

\newcommand{\dblfigrule}{\vspace*{-1pt}%
\noindent{\color{cream}\rule[-\figrulesep]{\textwidth}{1.5pt}} }

\makeatother

\twocolumn[
  \begin{@twocolumnfalse}
\vspace{1em}
\sffamily
\begin{tabular}{m{4.5cm} p{13.5cm} }

\includegraphics{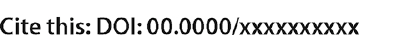} & \noindent\LARGE{\textbf{Exploring particle dynamics in flowing complex fluids \mbox{using} differential dynamic microscopy}} \\
\vspace{0.3cm} & \vspace{0.3cm} \\

& \noindent\large{James A. Richards$^{\ast}$, Vincent A. Martinez and Jochen Arlt}\\
\includegraphics{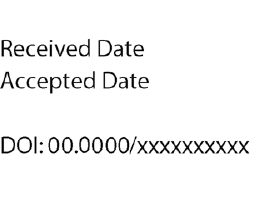} & \noindent\normalsize{Microscopic dynamics reveal the origin of the bulk rheological response in complex fluids. In model systems particle motion can be tracked, but for industrially relevant samples this is often impossible. Here we adapt differential dynamic microscopy (DDM) to study flowing highly-concentrated samples without particle resolution. By combining an investigation of oscillatory flow, using a novel ``echo-DDM'' analysis, and steady shear, through flow-DDM, we characterise the yielding of a silicone oil emulsion on both the microscopic and bulk level. Through measuring the rate of shear-induced droplet rearrangements and the flow velocity, the transition from a solid-like to liquid-like state is shown to occur in two steps: with droplet mobilisation marking the limit of linear visco-elasticity, followed by the development of shear localisation and macroscopic yielding. Using this suite of techniques, such insight could be developed for a wide variety of challenging complex fluids.}
\end{tabular}
\end{@twocolumnfalse} \vspace{0.6cm}
]
\renewcommand*\rmdefault{bch}\normalfont\upshape
\rmfamily
\section*{}
\vspace{-1cm}

\footnotetext{\textit{Edinburgh Complex Fluids Partnership and School of Physics and Astronomy, James Clerk Maxwell Building, Peter Guthrie Tait Road, King's Buildings, Edinburgh, United Kingdom, EH9 3FD. E-mail: james.a.richards@ed.ac.uk}}
\footnotetext{\dag~Electronic Supplementary Information (ESI) available: containing parallel sector echo-DDM comparison, echo-DDM ISF fitting details, fitting details for flow-DDM and imaging of diluted emulsions.}

\section{Introduction\label{sec:intro}}

Throughout the life-cycle of many materials they must flow as a multi-phase complex fluid. This may happen during processing, for example in making chocolate\cite{blanco2019conching} and ceramics;\cite{mbarki2017linking} transport, \textit{e.g.}, pumping mine tailings;\cite{boger2013rheology} or even in their application, such as consumer formulations like skin creams.\cite{kwak2015rheological} At all of these stages it is essential to understand and control how they respond to the applied shear stresses and strains. Using our previous examples, this is to ensure materials form the correct shape, do not block pipes and give the correct sensory properties. In each case the complex fluid is often highly concentrated, where the phase dispersed in a continuous background solvent is at a high volume fraction. The dispersed phase may vary, from the colloidal scale ($\SI{10}{\nano\metre}\lesssim$ diameter, $d \lesssim \SI{1}{\micro\metre}$) into the granular ($d \gg \SI{1}{\micro\metre}$), with the ``particles'' composed of a solid, immiscible liquid droplet, gas bubble, particle aggregate, or even a swollen polymeric gel.\cite{bonn2017yield}

The high volume fraction of this dispersed phase can arise from a desire for efficiency, as in reducing the water used in mine tailings\cite{boger2013rheology} or cocoa butter in chocolate.\cite{blanco2019conching} Alternatively, it may be to give desirable rheological properties, as at high volume fractions the system may become arrested and able to support a static stress like a solid. Returning to our examples, this would enable the system to retain a formed shape,\cite{mbarki2017linking} prevent sedimentation of suspended particles,\cite{ovarlez2012shear} or remain on the skin after dispensing.\cite{kwak2015rheological} Arrest may happen through various mechanisms: a glass transition;\cite{hunter2012physics} soft-jamming where particles are compressed;\cite{liu1998jamming} and shear jamming, with a compressive network.\cite{cates1998jamming} The links between these various states remains an active area of investigation.\cite{mari2009jamming}

These arrested materials are often termed ``soft-solids'', as under moderate stress ($\lesssim \SI{D3}{\pascal}$) the system may begin to flow in a liquid-like state. Understanding this yielding transition is vital for controlling a myriad of fluids. The challenge is one ubiquitous to soft matter as it involves bridging lengthscales, a theme found from controlling viral infections\cite{poon2020soft} to creating bio-mimetic structural colours.\cite{sicher2021structural} The particular challenge in understanding yielding is relating particle-particle interactions to the bulk rheology. At the local level, yielding involves cooperative rearrangements of several particles. When the system is sheared the particles must move around one another to accommodate the applied strain. These shear-induced particle rearrangements are non-affine motion in addition to the affine shear, \textit{i.e.}~the bulk flow.\cite{utter2008experimental} As this motion results from many particle interactions, much like the thermal motion in a liquid that causes Brownian motion, it is often treated as diffusive in nature.\cite{artoni2021self}

Much progress has been made in understanding yielding in well-controlled model systems by simultaneously looking at the local particle rearrangements using microscopy and measuring the bulk rheology. This has been achieved in soft-jammed systems with a silicone oil emulsion in a water-glycerol mixture. This system enables particle-level resolution at high magnification.\cite{knowlton2014microscopic,vasisht2018rate} Particle rearrangements have been probed when applying increasing strain in oscillatory shear, to look at the onset of yielding, and also with a continuous shear rate, to see how rearrangement types depend on the rate of bulk deformation. However, such microscopic insights are currently restricted to well-controlled model systems and these simple systems can be challenging to link back to industrially relevant materials. This limitation is due to the need for particle-level resolution. 

To enable looking at a wider variety of systems, we show how differential dynamic microscopy (DDM), an image analysis method for looking at the microscopic dynamics of quiescent complex fluids without particle resolution,\cite{cerbino2008differential} can be adapted to probe local particle rearrangements under shear. In particular, we develop "echo-DDM", a novel DDM analysis method to characterise the microscopic shear-induced particle rearrangements under oscillatory flow. We verify echo-DDM using a dilute colloidal suspension and then explore its potential using a silicone-oil emulsion imaged with rheo-confocal microscopy.\cite{besseling2009quantitative} Using a combination of echo-DDM and flow-DDM,\cite{richards2021particle} we investigate both oscillatory and steady shear, respectively, to fully characterise yielding of such a flowing complex fluid. This allows a comparison of the microscopic and bulk behaviour under both steady and oscillatory shear flows, which reveal a comprehensive view of the yielding process. Importantly, without the need for particle-level resolution many more particles can be monitored at the same time compared to particle-tracking. This results in rapid data acquisition,\cite{martinez2012differential} comparable in time to rheological tests, making the technique further suited for looking at various complex fluids.

\section{Differential dynamic microscopy\label{sec:DDM}}

Before looking at flowing systems, we shall consider the analysis of the microscopic dynamics of a quiescent system using differential dynamic microscopy. These dynamics could be, for example, diffusive Brownian motion\cite{cerbino2008differential} or micro-organism swimming.\cite{wilson2011differential,martinez2012differential,jepson2019high} Here, we highlight how complications arise in a flowing system and investigate how the effects of flow may be mitigated. Ultimately we present a novel analysis scheme that can robustly extract the microscopic shear-induced particle rearrangements, \textit{i.e.}~non-affine dynamics, in a flowing system and check that the effects of the affine flow do not dominate the measurement.

\subsection{Quiescent systems}

Differential dynamic microscopy\cite{cerbino2008differential} allows characterisation of the spatio-temporal density fluctuations within a sample by analysing the fluctuating intensity due to particle motion from microscopy movies. Specifically, one computes the differential intensity correlation function (DICF), also known as the image structure function: 
\begin{equation}
    g(\vec q, \tau) = \langle |\tilde{I}(\vec q,t + \tau) - \tilde{I} (\vec q, t)|^2 \rangle_t,
    \label{eq:DICF}
\end{equation}
with $\tilde{I} (\vec{q},t)$ the Fourier transform of the intensity, $I(\vec{r},t)$, at pixel position, $\vec{r}$, and time, $t$; $\tau$ is the delay time. Under appropriate imaging conditions and assuming the intensity fluctuations are proportional to density fluctuations, the DICF can be related to the real part of the intermediate scattering function (ISF), $f(\vec{q},\tau)$:\cite{wilson2011differential}
\begin{equation}
    g(\vec q, \tau) = A(\vec q)\left\{1 - \Re\left[ f(\vec q,\tau)\right] \right\} + B(\vec q).
    \label{eq:DICFfit}
\end{equation}
The signal amplitude, $A(\vec q)$, depends on sample properties and the imaging system, and $B(\vec q)$ is the instrumental noise. The ISF is a function of the particle displacement $\delta\vec r$: 
\begin{equation}
    f(\vec q, \tau) = \left \langle e^{i\vec q \cdot \delta \vec r_j} \right \rangle_{t,j},
    \label{eq:ISF_explicit}
\end{equation}
with brackets denoting averages over all particles $j$ and time $t$, and $\delta\vec r_j=\vec r_j(t+\tau) - \vec r_j(t)$ is the displacement of particle $j$ between $t$ and $t+\tau$. Isotropic motion allows radial averaging, so that $f(q,\tau)=\left<f(\vec{q},\tau)\right>_{|\vec{q}|=q}$. Fitting the DICF with a parametrised ISF allows extraction of key dynamical quantities. For non-interacting Brownian particles, such as a dilute colloidal suspension, the free diffusion coefficient ($D_0$) can be measured via
\begin{equation}
    f_D = \exp\left(-D_0q^2\tau\right).
    \label{eq:f_diff}
\end{equation}
For interacting Brownian particles, \textit{i.e.}~a concentrated colloidal suspension, deviations from this simple exponential, due to particle interactions, are usually captured through a stretched exponential, $f_D=\exp[-(q^2D\tau)^\beta]$, with $\beta<1$ the stretch exponent.\cite{martinez2011dynamics} For very high concentrations, typical in colloidal glasses, the ISF may no longer decay to zero, suggesting the existence of  `fixed' particles that do not rearrange, often characterised by the non-ergodicity parameter.\cite{martinez2011dynamics}

\subsection{Flowing systems}

Under shear, the motion of particles has three components. In addition to the quiescent microscopic dynamics, particles have correlated motion due to the bulk flow and extra microscopic dynamics due to local shear-induced rearrangements. For the latter, we assume a diffusive form,\cite{khabaz2020particle} isotropic for simplicity, such that the particle motion is separated into microscopic diffusive-like motion and bulk flow: $\delta \vec r = \delta \vec r_D + \delta \vec r_{\rm bulk}$. In contrast to the non-affine motion, the displacement due to the applied flow will be deterministic (although for complex fluids it may not be known \textit{a priori}) and direction dependent. The ISF can therefore be split into two components,\cite{aime2019probing}
\begin{equation}
    f(\vec q,\tau) = f_D(|q|,\tau) \cdot f_{\rm bulk}(\vec q,\tau),
    \label{eq:f_product}
\end{equation}
where $f_{\rm bulk}$ is anisotropic. This decoupling means that if the system decorrelates due to affine motion, $f_{\rm bulk}=0$, prior to decorrelation due to diffusive motion, then $f_D$ cannot be characterised from the total ISF (as $f = f_D\cdot f_{\rm bulk} = 0$ contains no information on the microscopic diffusive dynamics). We must therefore focus on ensuring $f_{\rm bulk} \approx 1$. Additionally, as DDM utilises a temporal average over $t$ (Eq.~\ref{eq:DICF}), we restrict ourselves to rheometric protocols that measure stationary properties: oscillatory shear (storage and loss moduli) and steady shear (stress as a function of shear rate, $\sigma(\dot\gamma)$). We use the subscripts ``flow'' and ``osc'' to refer to steady and oscillatory shear, respectively.

\subsubsection{Steady shear and flow-DDM\label{sec:theory_flow}}

In steady flows, particles are subject to diffusive dynamics and to advection across the field of view with a fixed velocity, $\vec v$. To characterise the non-affine motion of the particles, we must effectively reduce the impact of the affine motion due to the advective flow. This is obtained using flow-DDM,\cite{richards2021particle} which we have introduced elsewhere and we present only a brief summary. Crucially, flow-DDM performs DDM analysis in the co-moving frame by calculating the flow-corrected DICF, $\bar{g}$,
\begin{equation}
    \begin{split}
    \bar g(\vec q, \tau) &= \langle |\tilde I(\vec q,t + \tau)e^{-i\vec q \cdot \vec v \tau} - \tilde I (\vec q, t)|^2 \rangle_t \\
    &= A(\vec q)\left[1 - \bar f(\vec q,\tau) \right] + B(\vec q).
    \end{split} \label{eq:DICFcorr}
\end{equation}
Equation~\ref{eq:DICFcorr} requires knowledge of $\vec{v}$, which we measure using $\varphi$-DM,\cite{colin2014fast} a Fourier-space method that can be efficiently combined with DDM.\cite{richards2021particle}

To further minimise the impact of flow, we calculate $\bar g ^{\perp}$ by averaging $\bar g$ over a narrow range of $\vec q$ in a sector perpendicular to the flow direction ($\pm \theta = \SI{3}{\degree}$). This reduces the impact of any residual distribution in flow speeds, $\pm \Delta v$, due to, \textit{e.g.}~the shear rate across the optical section, as the component in this direction is small. The microscopic dynamics can then be extracted from fitting $\bar g ^{\perp}$, using Eqs.~\ref{eq:DICFfit} and \ref{eq:f_diff}, assuming $\bar{f}^{\perp}$ is mainly dominated by diffusive dynamics. To check that diffusive dynamics indeed dominate the decorrelation of $\bar g ^{\perp}$, we compare the perpendicular sector to an adjacent sector (that is slightly closer to the flow direction), see ESI$^{\dag}$ Sec.~S1. If the residual velocity distribution dominates over diffusive dynamics, then the decorrelation time will depend on sector.\cite{richards2021particle} We find in all results presented below that the microscopic dynamics dominate decorrelation in $\bar g ^{\perp}$ and $\bar f ^{\perp}$ is indeed mainly related to non-affine motion.

\subsubsection{Oscillatory flow and echo-DDM \label{sec:echo}}

In contrast to steady shear, oscillatory flow varies strongly with time as a sinusoidal strain is applied to the sample, $\gamma(t) = \gamma_0 \sin(\omega t)$, with a strain amplitude ($\gamma_0$) and angular frequency ($\omega = 2\pi \times$ frequency in Hz). At a depth $h$ into the sample, the particle displacement due to oscillatory flow along $\hat x$ is then,
\begin{equation}
    \delta \vec r_{\rm osc}(\tau) = h\gamma_0\hat x \{\sin[\omega(t+\tau)]-\sin(\omega t)\}.\footnote{For didactic purposes we assume no slip [$v(h\!=\!0)=0$] and a uniform velocity gradient. However, the following results hold without these conditions, as may occur for complex fluids.}
\end{equation}
The changing speed and direction of flow inherent to oscillatory shear makes the quantitative determination of $\vec v(t)$ challenging (see Sec.~\ref{sec:results} and thus the application of flow-DDM). Instead, we show below that the periodic return of the sample to the original bulk position can be exploited to extract local shear-induced rearrangements.

If an image is taken at $\tau = 2\pi/\omega$, $\delta r_{\rm osc} = 0$ and $f_{\rm osc} \approx \exp(i\vec q \cdot \vec 0) \approx 1$. ``Stroboscopic imaging'' requires timing to the displacement either with triggering or even pausing flow.\cite{knowlton2014microscopic} However, this allows only one image per cycle and requires integration of imaging and shear that restricts widespread adoption. If instead a continuous movie is taken at a frame rate ($t_f^{-1}$) higher than the oscillation frequency ($\omega/2\pi$) we can guarantee images \emph{close} to a cycle apart, $\pm t_f$, and record many images per cycle for faster data acquisition. However, if the oscillation period and imaging frequency are not precisely synchronised this will cause an offset in the images, which will be greater as $\gamma_0 h$ is increased. To maximise the depths and strain amplitudes that can be probed (taking $t_f$ and $\omega$ as fixed) we must minimise the impact of this offset, such that $f_{\rm flow} \approx 1$.

\begin{figure}
    \centering
    \includegraphics{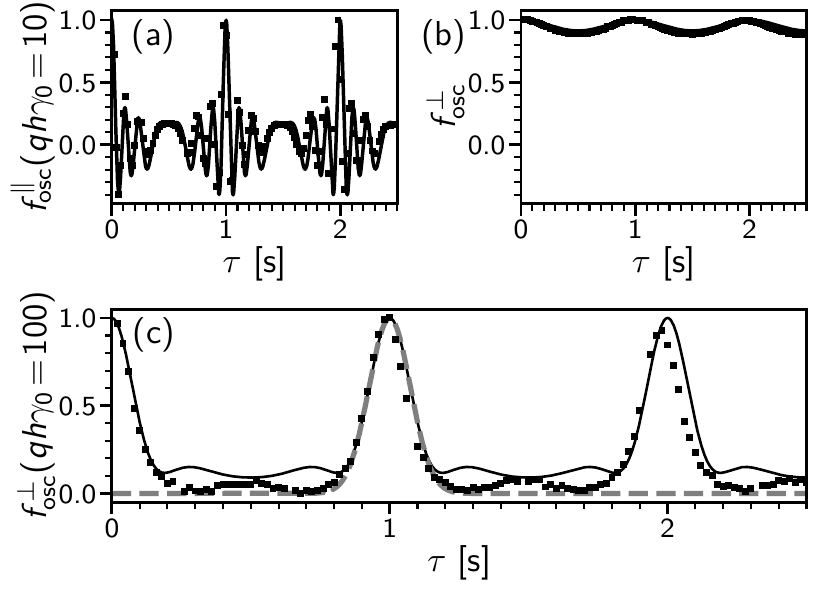}
    \caption{Oscillatory flow of a dynamics-free sample. (a) Parallel ISF component, $f_{\rm osc}^{\parallel}(\tau)$, with delay time, $\tau$, for moderate displacement at \SI{1}{\hertz}. Lines, numerical solution of Eq.~\ref{eq:ISF_osc_general}. Symbols, data for colloids in high-viscosity solvent (90~wt\% glycerol-water, see Sec.~\ref{sec:materials} for details) at $\gamma_0 = 0.5$, $h = \SI{10}{\micro\metre}$ and $q = \SI{2}{\per\micro\metre}$. Sector width, $\theta = \SI{3}{\degree}$, $\rightarrow qh\gamma_0\theta = 0.5$. (b)~Perpendicular sector, $f_{\rm osc}^{\perp}(\tau)$, corresponding to (b). (c)~Perpendicular sector at high displacement, $qh\gamma_0\theta = 5$. Lines: bold (black), numerical solution to Eq.~\ref{eq:ISF_osc_general}; dashed, Gaussian approximation (Eq.~\ref{eq:f_perp_high}). Symbols from data at $\gamma_0 = 2$, $h = \SI{10}{\micro\metre}$ and $q = \SI{5}{\per\micro\metre}$.}
    \label{fig:osc_ddm}
\end{figure}

As the offset from frame timing will be in the flow direction, averaging $g$ over a narrow $\vec{q}$ range in a sector perpendicular to the flow will minimise the impact flow, just as with flow-DDM. This is confirmed by numerically calculating
\begin{equation}
    f_{\rm osc}(\vec q, \tau) = \frac{1}{T} \int_0^{T} \cos\left[\vec q \cdot \hat x \delta r_{\rm osc}(t,\tau) \right] {\rm d}t,
    \label{eq:ISF_osc_general}
\end{equation}
in a sector parallel to the flow, Fig.~\ref{fig:osc_ddm}(a), and perpendicular, (b), at the same oscillation amplitude. At small displacement amplitudes ($\gamma_0 h$), a sharply peaked, rapid oscillation in $f^{\parallel}_{\rm osc}$ is reduced to a weak modulation in $f^{\perp}_{\rm osc}$ at the oscillation frequency. Sampling at a finite frequency, the peak (or ``echo'') value where the sample returns to its original bulk position ($f_{\rm osc} = 1$) is recovered in the perpendicular sector, but may be missed in the parallel sector even at small oscillation amplitudes. The peaks in $f^{\perp}_{\rm osc}$ correspond to minima in $g^{\perp}_{\rm osc}$. Locating minima, rather than taking $\tau = \mathbb{Z}T$, allows for any shift in timing between applied shear and imaging over the course of the experiment. The microscopic dynamics can be measured in $g^{\perp}_{\rm echo}$, constructed from these minima. We term construction and analysis of $g^{\perp}_{\rm echo}$, ``echo-DDM''.

We now consider the limit of echo-DDM, where the impact of the flow velocity can no longer be neglected. About $\tau \approx \mathbb{Z}T$, we can series expand Eq.~\ref{eq:ISF_osc_general} so that $\delta r_{\rm flow}(\tau,t) \approx h\gamma_0 \omega (\tau - \mathbb{Z} T) \cos(\omega t)$. The phase shift in Fourier space is then different for each $t$, as the instantaneous velocities vary sinusoidally with $t$. Away from $\tau - \mathbb{Z} T = 0$ there is then a summation of many different phase shifts in Eq.~\ref{eq:ISF_osc_general} that will lead to rapid decorrelation. This results in a sharp peak in $f^{\perp}_{\rm osc}$, Fig.~\ref{fig:osc_ddm}(c), with the width set by the oscillation amplitude. We can quantitatively approximate this peak with a Gaussian form:
\begin{equation}
    f^{\perp}_{\rm osc}(q\theta h \gamma_0 \gg 1, \tau \!-\! \mathbb{Z}T \ll 1 ) \approx \exp \left [ - \frac{1}{2} \left( \frac{q\theta\gamma_0 h
    \tau\omega}{2.45}\right)^2 \right].
    \label{eq:f_perp_high}
\end{equation}

The width of the peak in $f_{\rm osc}$ decreases with increasing oscillation amplitude. With an imaging rate $t_f^{-1}$, this sets a practical limit for echo-DDM, as the peak must be wide enough for identification of the maximum value and recovery of $f_{\rm osc} \approx 1$. For $\pm t_f$ to be within 95\% of the maximum value, we require $q\theta h \gamma_0 t_f/T < 0.08$, based on the form of the Gaussian peak. Imaging with higher $T/t_f$ would result in better stroboscopic timing, however these variables are typically fixed by the imaging system ($t_f$) and the rheology ($T$). For a fixed $t_f/T$, this requirement sets a maximum $q \gamma_0 h < 80$ (using our set experimental values of $\theta = \SI{3}{\degree}$, $t_f = \SI{0.02}{\second}$, $T = \SI{1}{\second}$). For a given lengthscale, \textit{e.g.}~set by the fluid structure, this sets a limit for the displacement amplitude, $\gamma_0 h$. For example, at $q = \SI{3.5}{\per\micro\metre}$ (see Sec.~\ref{sec:results}), this corresponds to an amplitude of $\pm \SI{23}{\micro\metre}$. In complex fluids this quantity may not be known, but the peak width still gives information on the flow velocity and whether the impact of flow can be neglected. This protocol therefore meets the aims of reducing the impact of oscillatory flow and determining reliability.

\section{Materials and Methods\label{sec:materials}}

\subsection{Dilute colloidal suspensions}

Looking at strain-dependent dynamics creates a particular issue. With increasing deformation the microscopic dynamics are predicted to speed up, however the residual effects of affine deformation measured through both flow-DDM and echo-DDM will also increase. In this respect, we have previously established the limits of flow-DDM.\cite{richards2021particle} In Sec.~\ref{sec:dilute}, we validate echo-DDM and investigate its limits using a system whose dynamics do not depend on deformation, \textit{i.e.}~a dilute colloidal suspension with a well-defined diffusion coefficient.

In echo-DDM the dynamics are probed on time scales corresponding to multiple oscillation cycles, $\tau > T$. In an arrested, highly concentrated system the quiescent dynamics are suppressed by particle interactions and the rate of dynamics is intrinsically coupled to the applied strain that sets our temporal resolution. However, in a dilute system the diffusive Brownian motion could cause decorrelation within a single cycle (Eq.~\ref{eq:f_diff}, $D_0 q^2 \tau \gg 1$). To prevent this and allow echo-DDM to be tested on a dilute colloidal suspension a small diffusion coefficient is required. When applying a \SI{1}{\hertz} oscillation frequency, a Brownian time of multiple seconds is needed. To reach this criterion with a micron-sized system, suspensions of 0.5\% volume fraction were prepared using \SI{2.4}{\micro\metre}\cite{richards2021particle} poly-(vinyl pyrrolidone)--stabilised fluorescein-dyed poly-(methyl methacrylate) particles in a 60~wt\% glycerol-water mixture ($\eta_s = \SI{12}{\milli\pascal\second} \rightarrow D = \SI{0.014}{\micro\metre^2 \per\second}$). To screen electrostatic interactions \SI{0.1}{\molar} sodium chloride was added. To create a system with negligible dynamics a 90~wt\% glycerol-water mixture ($\eta_s = \SI{200}{\milli\pascal\second}$) was also used.

Suspensions were sheared in a rheo-confocal imaging set-up\cite{besseling2009quantitative} that couples an inverted laser-scanning confocal microscope (Leica SP8) to a stress-controlled rheometer (Anton-Paar MCR 301, \SI{50}{\milli\metre} \SI{1}{\degree} cone-plate geometry), Fig.~\ref{fig:confocal}(a). Samples were imaged at a depth of \SI{10}{\micro\metre} through a glass coverslip. A 20x/0.75 objective was used to image a $\SI{466}{\micro\metre}\times \SI{116}{\micro\metre}$ region (1024~px $\times$ 256~px), Fig.~\ref{fig:confocal}(b). Movies were recorded at 50 frames per second for \SI{200}{\second} of continuous oscillation at a given strain amplitude from $\gamma_0 = 1\%$ to 500\%. The DICF was calculated using windowing to reduce edge effects.\cite{giavazzi2017image} Over many oscillation cycles a slow drift of the sample in the radial direction ($\sim$ perpendicular to the applied flow) was observed due to thermal expansion or contraction of the gap between cone and plate. The DICF was therefore calculated with a correction for the mean velocity due to radial drift.\footnote{Radial drift correction using Eq.~\ref{eq:DICFcorr} and a time-independent $\vec v$ measured over multiple oscillation cycles leaves the applied oscillatory shear. As the radial drift is a weak experimental artefact we do not refer explicitly to $\bar g$, as with steady shear.}

\subsection{Concentrated silicone oil emulsions}

\begin{figure}[tb]
    \centering
    \includegraphics[width=\columnwidth]{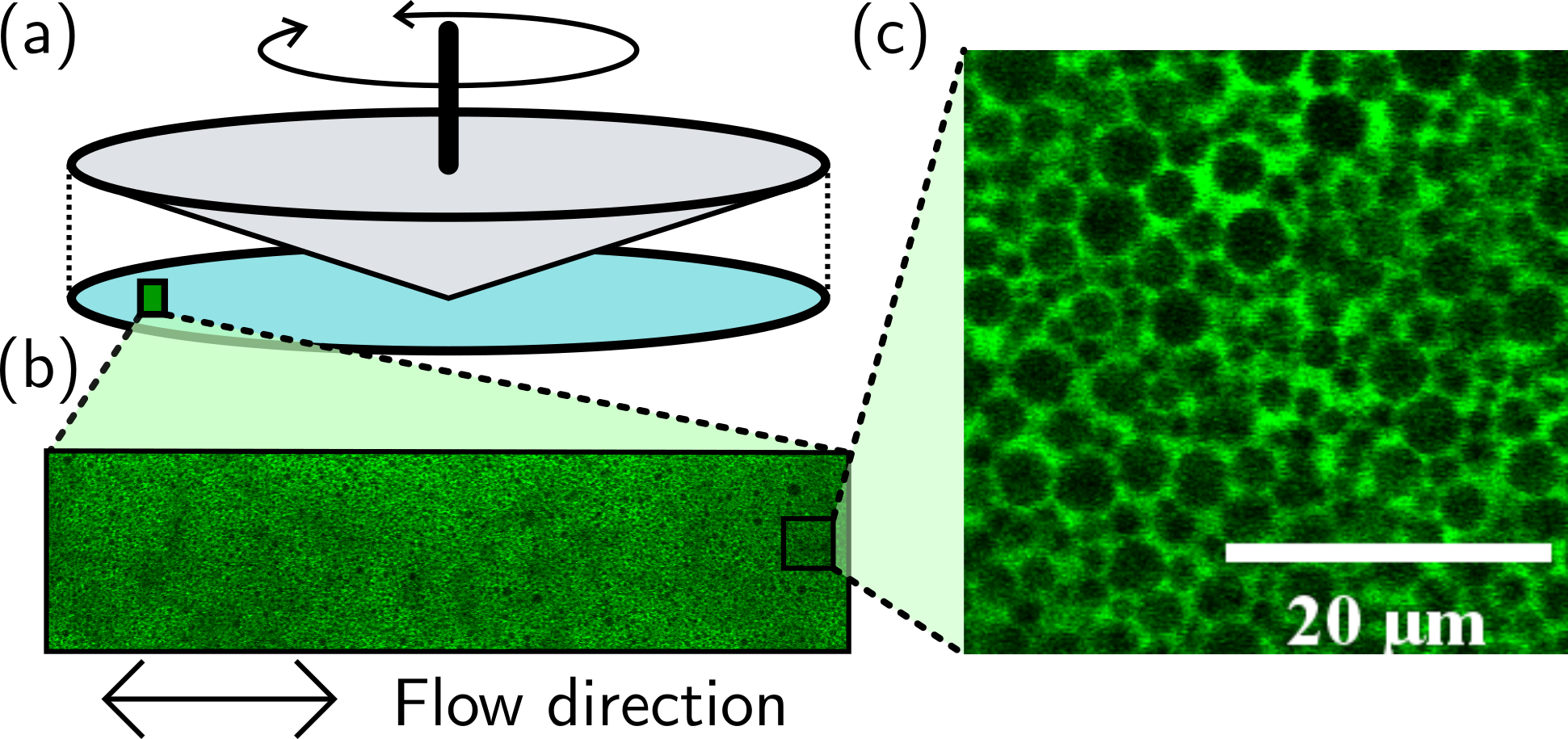}
    \caption{Rheo-confocal imaging of an emulsion. (a) Cone-plate rheometric flow, imaging through coverslip. (b)~Region imaged for DDM processing with 20$\times$ objective. (c)~High-magnification image (63$\times$), see scale bar and equivalent area shown in (b)}
    \label{fig:confocal}
\end{figure}

For probing the non-linear microscopic dynamics of complex fluid, we used a highly concentrated silicone-oil emulsion. A base emulsion was prepared from silicone oil (viscosity 50~cSt, Sigma Aldrich) and a \SI{0.1}{\molar} surfactant solution of sodium dodecyl sulphate (SDS). The polydisperse oil-in-water emulsion was fractionated.\cite{bibette1991depletion} This removes large droplets (creaming at $<\SI{30}{\milli\molar}$ SDS) that have been shown to influence microscopic rearrangements under shear\cite{clara2015affine} and sub-micron droplets (remaining suspended at $>\SI{30}{\milli\molar}$ SDS). The fractionated emulsion was transferred to an index-matched 53~wt\% glycerol-water mixture with \SI{0.2}{\milli\molar} fluorescein sodium salt (Sigma Aldrich), such that depletion-induced attraction is negligible (SDS concentration below critical micellar concentration\cite{becu2006yielding}). The final volume fraction, $\phi = 0.69$, was reached from centrifugation at $\sim 800$g. Under centrifugal separation the system becomes jammed and particles are elastically compressed into each other above random close packing, $\phi_{\rm rcp}$. The resulting sample is shown under high-magnification (63x) in Fig.~\ref{fig:confocal}(c), consisting of $\sim$\SIrange{1}{3}{\micro\metre} droplets. For monodisperse spheres $\phi_{\rm rcp}= 0.64 < \phi = 0.69$,\cite{bernal1960packing} although as $\phi_{\rm rcp}$ is increased by polydispersity, $\phi- \phi_{\rm rcp}$ is not precisely known.\cite{desmond2014influence} 

For rheo-confocal imaging a roughened coverslip (roughness $\approx \SI{0.9}{\micro\metre}$) was used to prevent slip.\cite{meeker2004slip} In oscillatory shear, amplitudes of 2.5\% to 160\% were applied at \SI{1.0}{\hertz}. To reduce the impact of photobleaching, the sample was pre-sheared at $\dot\gamma = \SI{1.0}{\per\second}$ for \SI{60}{\second} before each applied $\gamma_0$. Movies were recorded from the start of the first oscillation for 180 cycles. For velocimetry, movies were also recorded at \SI{0.1}{\hertz} over a smaller number of cycles (30) with $\gamma_0 = $\SIrange{2.5}{400}{\%}. Under steady shear, movies were taken \SI{20}{\second} from the start of shear with the frame rate adjusted to the shear rate, from 10 to 85~fps, see Fig.~\ref{fig:steady} caption for details. At the highest frame rate a reduced 1024$\times$128 image was recorded (with $\theta = \SI{5}{\degree}$ then used in defining $\bar g^{\perp}$). The shear-induced dynamics were then extracted from the movies by using a combination of echo-DDM and flow-DDM using Eqs.~\ref{eq:DICFfit} and \ref{eq:f_diff}.

Experimentally, the bulk response was measured with a roughened \SI{40}{\milli\metre} diameter \SI{1}{\degree} cone-plate geometry (TA Instruments ARES-G2). For oscillatory rheology at \SI{1.0}{\hertz} the strain amplitude was varied logarithmically at 10 pts./decade from 0.02\% to 200\%. We report the average of an upsweep and downsweep of $\gamma_0$. For steady flow, we report the average of three separately loaded samples applying $\dot\gamma = $\SIrange{0.02}{2000}{\per\second} at 10 pts./decade. At each point the sample was equilibrated for \SI{10}{\second} and the response measured for \SI{20}{\second}. The reversibility of the response was ensured for each sample up to \SI{200}{\per\second} before each reported set of results.

\section{Results\label{sec:results}}

\subsection{Echo-DDM of a dilute suspension\label{sec:dilute}}

\begin{figure}[tb]
    \centering
    \includegraphics{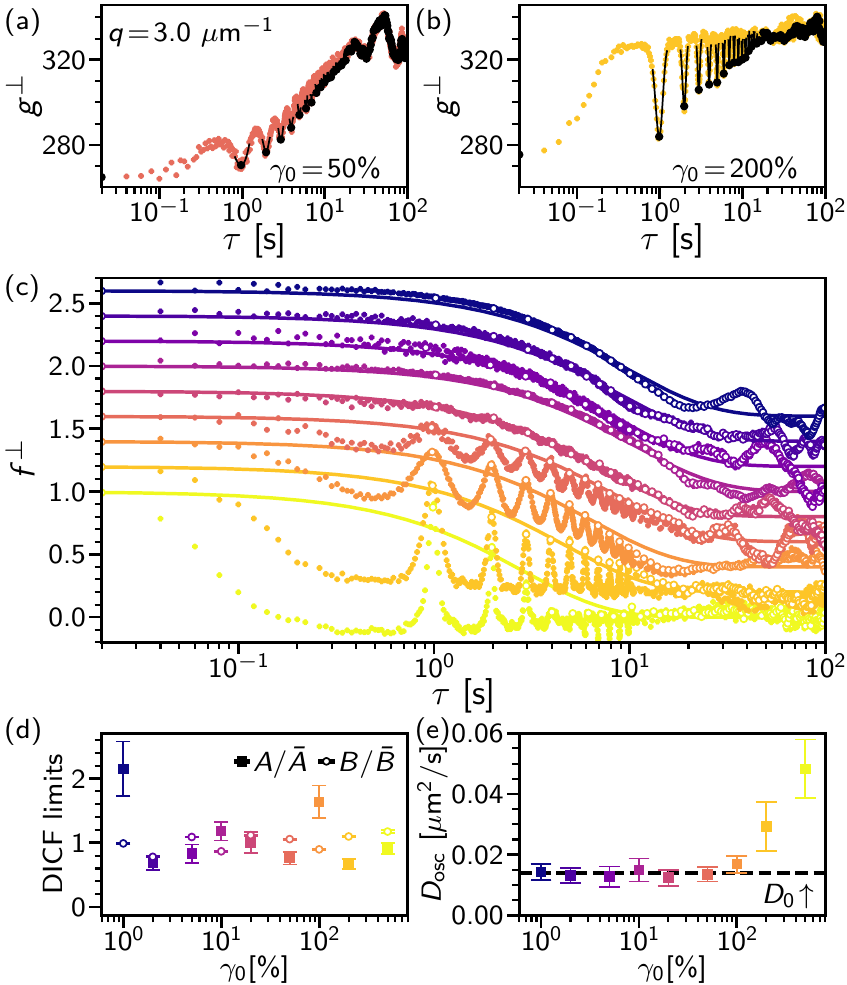}
    \caption{Echo-DDM of a dilute colloidal suspension. (a)~DICF in perpendicular sector, $g^{\perp}(\tau)$. at $q=\SI{3}{\micro\metre^{-1}}$ and applied strain amplitude, $\gamma_0 = 50\%$. Symbols: open (pink), recorded data; filled (black), $g^{\perp}_{\rm echo}$, for echo-DDM from minima fitted using Gaussian (line). (b)~Corresponding DICF and fitting at $\gamma_0 = 200\%$. (c)~Reconstructed ISF, $f^{\perp}(\tau)$ at $q=\SI{3}{\per\micro\metre}$, based on fitting to minima values in DICF. For clarity ISF shifted vertically with decreasing strain amplitude, $\gamma_0$. Symbols: dark (purple) to light (yellow) increasing $\gamma_0$, from 1 to 500\% [see (d) for $\gamma_0$]; open, values of $f^{\perp}_{\rm echo}$; and small filled, full ISF, $f^{\perp}$. Lines, diffusive ISF, $f_D$. (d)~DICF parameters. Symbols: filled, signal strength with strain [$A(q)$] normalised by average over $\gamma_0$, mean and error taken from $q = $~\SIrange{2.0}{3.5}{\per\micro\metre}; and open, equivalent for $B(q)$. (e)~Extracted diffusion coefficient with increasing $\gamma_0$, averaging over $q = $~\SIrange{2.0}{3.5}{\per\micro\metre}, compared to $D_0 = \SI{0.014}{\micro\metre^2\per\second}$(dashed line)}
    \label{fig:60glyc}
\end{figure}
Here, we validate echo-DDM and investigate its limits related to the residual effects of flow. In a system with negligible dynamics subjected to oscillatory flow only the impact of flow will be measured, \textit{i.e.}~$f_{\rm osc}$. This is realised using a dilute colloidal suspension in a highly viscous background solvent, 90~wt\% glycerol-water, to suppress diffusion ($D \approx \SI{D-3}{\micro\metre^2\per\second}$, so $f_{D} \approx 1$). Even at moderate oscillatory amplitudes, $\gamma_0 = 50\%$, the extracted ISF in the direction of shear ($f_{\rm osc}^{\parallel}$) is strongly impacted by the flow, Fig.~\ref{fig:osc_ddm}(a). About each oscillation period, where the particles return to their original position, $f_{\rm osc}^{\parallel}$ is rapidly oscillating and sharply peaked: recovering $f_{\rm osc} = 1$ would require precise stroboscopic timing. In contrast, the ISF in the perpendicular sector is only weakly modulated at the oscillation frequency, Fig.~\ref{fig:osc_ddm}(b): recovering $f_{\rm osc}^{\perp} = 1$ can be readily achieved. A broad peak remains in the perpendicular sector even when the impact of oscillatory flow is increased at a higher strain amplitude, $\gamma_ 0 =200\%$ in Fig.~\ref{fig:osc_ddm}(c). However, it should be stressed that there are experimental limits to echo-DDM due to the rheo-imaging geometry, as without careful measures over long times (say $\tau \gtrsim \SI{100}{\second}$) the sample may evaporate, drift due to thermal expansion of the geometry, sediment or photobleach.

To test how microscopic dynamics can be recovered via echo-DDM in the presence of oscillatory flow, we look at the same colloidal particles but in a less viscous background (60~wt\% glycerol-water), such that diffusion occurs over multiple cycles. To recover $f_{\rm osc}^{\perp} = 1$ we fit a Gaussian peak to the DICF minima at each cycle delay time, $\mathbb{Z}T\pm T/5$, Fig.~\ref{fig:60glyc}(a) at $\gamma_0 = 50\%$ and (b) at 200\% (black lines). The reconstructed minima (black squares) then give the echo-DICF, $g^{\perp}_{\rm echo}$. Where no peak can be located the points are taken directly from $g(\tau=\mathbb{Z}T)$, for example where the system has diffusively decorrelated or the impact of flow is negligible. Through fitting the extracted $g^{\perp}_{\rm echo}$, the echo-ISF ($f^{\perp}_{\rm echo}$) and full ISF, including the impact of flow ($f^{\perp}$), can be reconstructed, Fig.~\ref{fig:60glyc}(c) (open and small symbols, respectively). We find $g^{\perp}_{\rm echo}$ can be well fitted assuming diffusive dynamics, \textit{i.e.} 
\begin{equation}
    g^{\perp}_{\rm echo} = A(1-f_D)+B,
    \label{eq:g_perp_diff}
\end{equation}
as the system decorrelates over $\mathcal{O}(10)$ cycles, enabling the short-time [$B(q)$] and long-time plateaus to be captured, Fig.~\ref{fig:60glyc}. The signal strength and noise are insensitive to the applied oscillatory flow, Fig.~\ref{fig:60glyc}(d), although they do show variation, possibly due to fluctuations in the small number of imaged particles in a narrow section of a dilute colloidal suspension.

The form of the full ISF shows that in the perpendicular sector the impact of flow is negligible until $\gamma_0 = 10\%$, with the data following a diffusive form (line). Above this strain, the full ISF is impacted by the flow with an oscillation at the applied frequency, even in the perpendicular sector. However, the diffusion coefficient, $D_{\rm osc}$, measured from $f^{\perp}_{\rm echo}$ remains constant until $\gamma_0 = 100\%$ and comparable to the predicted $D_0$, Fig.~\ref{fig:60glyc}(e). Only at $\gamma_0 \geq 200\%$ does $D_{\rm osc}$ increase as the peaks in $f^{\perp}$ become too narrow to reliably extract $f^{\perp}_{\rm echo}$. This occurs at the limit based upon a 95\% peak value within $T\pm t_f$, with affine displacements $\gamma_0 h = \SI{20}{\micro\metre}$, see Sec.~\ref{sec:echo}. We have therefore shown the effects of oscillatory flow on DDM, where the impact has been minimised by using a narrow sector perpendicular to the flow, the importance of which is highlighted by comparison with the sector in the flow direction (ESI$^{\dag}$ Sec.~S2). The microscopic diffusive dynamics can be recovered up to a limit based upon $q$, the displacement of the flow and the ratio of frame rate to oscillation frequency. Finally, the peak width in the full DICF gives a qualitative interpretation of the impact of flow, to establish whether the measurement of the microscopic dynamics is reliable.

\subsection{Concentrated silicone-oil emulsion}

Having established echo-DDM to measure microscopic dynamics in systems undergoing oscillatory shear, we now demonstrate that a combination of echo-DDM and flow-DDM allows a relation of microscopic dynamics to bulk rheology in a non-Newtonian complex fluid in both oscillatory and steady flow using a highly-concentrated silicone oil emulsion.

\subsubsection{Oscillatory flow\label{sec:osc_soe}}

\begin{figure}[tb]
    \centering
    \includegraphics{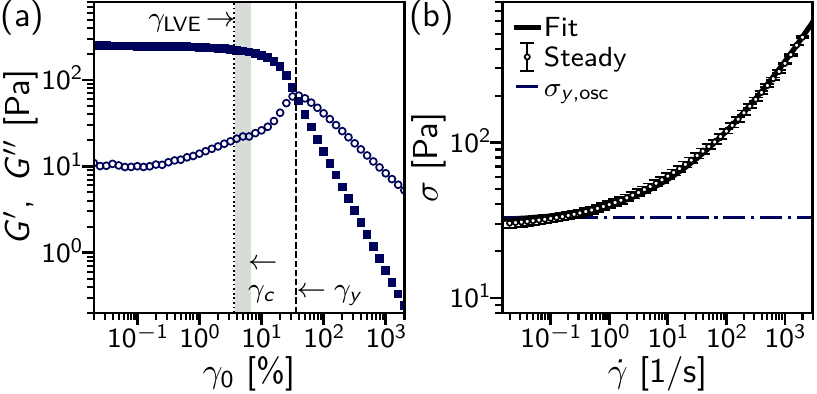}
    \caption{Rheological characterisation of a concentrated silicone-oil emulsion. (a)~Oscillatory rheology, storage ($G^{\prime}$, filled squares) and loss ($G^{\prime\prime}$, open circles) moduli vs strain amplitude, $\gamma_0$ at \SI{1.0}{\hertz}. Defined strains: dotted line, $\gamma_{\rm LVE} = 4\%$, end of linear elasticity at $G^{\prime} < 0.9G^{\prime}(\gamma_{0,\min}=0.02\%)$; shaded region, onset of microscopic rearrangements, $4.0 \% < \gamma_0 < 6.3\%$; dashed line, yield strain ($\gamma_y$) where $G^{\prime}=G^{\prime\prime}$. (b)~Steady-state rheology, stress vs shear rate, $\sigma(\dot\gamma)$. Symbols, data; solid line, three component model fit, see text; dashed line, yield stress defined from crossover of oscillatory moduli}
    \label{fig:rheo}
\end{figure}

With an applied oscillatory strain, at low $\gamma_0$ the mechanical behaviour of the emulsion is solid-like, with the storage modulus ($G^{\prime} \approx \SI{300}{\pascal}$) higher than the loss modulus ($G^{\prime\prime}$), Fig.~\ref{fig:rheo}(a). The stress is in phase with the applied strain. With increasing strain amplitude, after the initial plateau of the linear elastic regime ($\gamma_{\rm LVE} = 3.6\%$, dotted line) the emulsion begins to yield and transition to a fluid-like response: $G^{\prime}$ drops and $G^{\prime\prime}$ rises. At $\gamma_y = 36\%$ the moduli cross (dashed line), a point that is conventionally defined as yielding.\cite{dinkgreve2016different} At higher strains, with $G^{\prime}$ sharply dropping below $G^{\prime\prime}$, the stress response is liquid-like and in phase with the applied shear rate. The wide range of strain in this transition, from the moduli leaving the linear visco-elastic region to crossing-over, represents that macroscopic yielding is a gradual transition, which is not clearly defined from the bulk response alone. We show below that echo-DDM allows one to determine a far narrower range of strain for the yielding transition from the microscopic dynamics.

\begin{figure}[tb]
    \centering
    \includegraphics{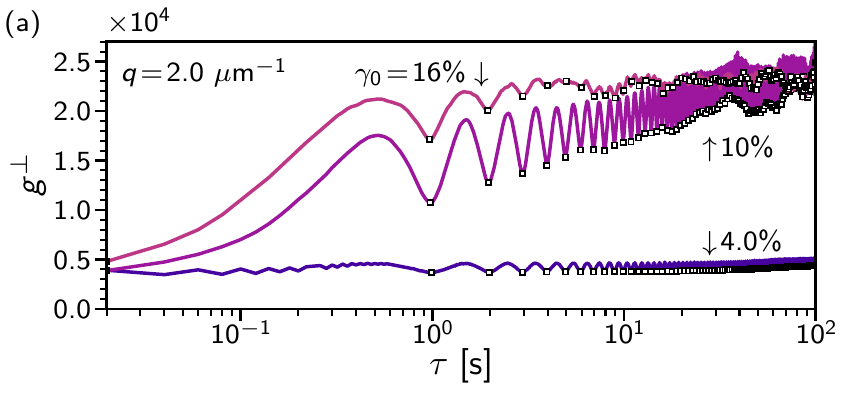}
    \includegraphics{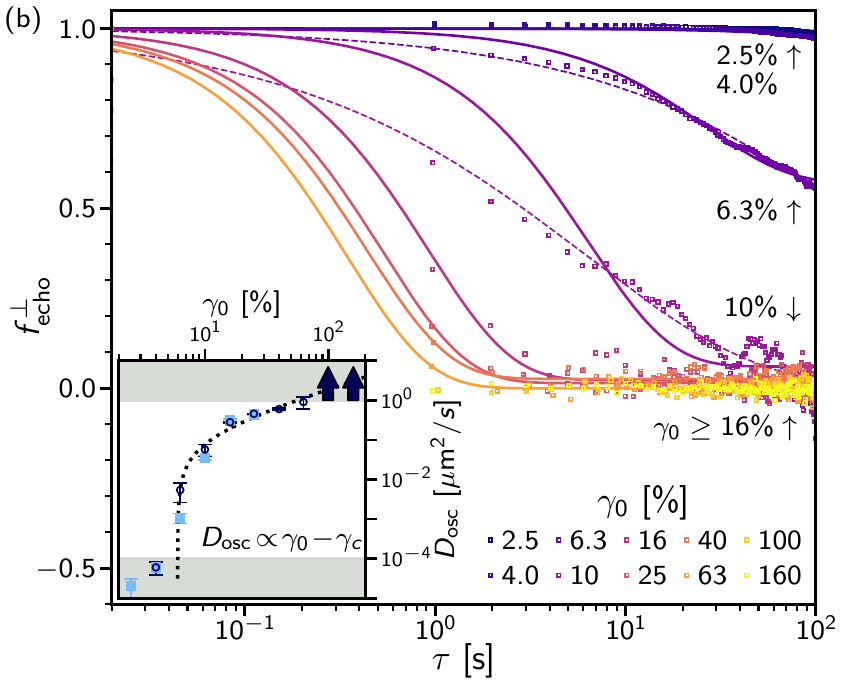}
    \caption{Microscopic dynamics for concentrated silicone-oil emulsion under oscillatory shear at \SI{1}{\hertz} with increasing strain amplitude. (a) DICF shown for perpendicular sector, $g^{\perp}(\tau)$ at $q = \SI{2}{\per\micro\metre}$. Lines: dark (indigo), $\gamma_0 = 4\%$; medium (purple), $10\%$; and light (violet), 16\%. Open symbols, extracted DICF values, $g^{\perp}_{\rm echo}$. (b)~Reconstructed ISF. Points, data for $q = \SI{2}{\per\micro\metre}$ for given $\gamma_0$ (see legend). Lines: solid, diffusive fit with varying proportion of mobile particles; and dotted, stretched exponential shown for $\gamma_0 = 6.3\%$ and 10\%. Inset: extracted diffusivities, $D_{\rm osc}$, with $\gamma_0$. Symbols: dark (blue) squares, varying $n_{\rm mob}$; light (blue) circles, stretched exponential. Shaded regions where diffusivity is too slow ($t_{\max}$ setting $D>\SI{D-4}{\micro\metre^2\per\second}$) or fast ($T$ setting $D<\SI{1}{\micro\metre^2\per\second}$. Dotted line, diffusivity proportional to irreversible strain rate, $D_{\rm osc} = (\gamma_0-\gamma_c) \times \SI{2}{\micro\metre^2\per\second}$. Average performed over $q = $\SIrange{2.0}{2.3}{\per\micro\metre}, error from standard deviation}
    \label{fig:soe}
\end{figure}

On the microscopic level, using echo-DDM we probe a range of strains from the linear visco-elastic region ($\gamma_0 = 2.5\%$) to above the bulk yield point, $\gamma_0 > 40\%$. The full DICF in the perpendicular sector, $g^{\perp}$, is shown for several strain amplitudes, all below the macroscopic yield point, $\gamma_0 = 4\%$, 10\% and $16\%$, Fig.~\ref{fig:soe}(a)~(lines). The microscopic dynamics are again isolated by reconstructing the minima of $g^{\perp}$ at each oscillation cycle to yield $g^{\perp}_{\rm echo}$ (open points). In contrast to a dilute colloidal suspension, the microscopic dynamics depend strongly on the applied strain amplitude. At $\gamma_0=4\%$, $g^{\perp}_{\rm echo}$ is near constant and the system remains correlated over the length of the experiment [dark (indigo)]. At $\gamma_0 = 10\%$, $g^{\perp}_{\rm echo}$ increases from the first cycle and continues to increase over multiple cycles as the system decorrelates and the droplets rearrange [medium (purple)]. However, at this strain amplitude no clear long--delay-time limit can be established. As the strain amplitude increases further, $\gamma_0 = 16\%$, the dynamics speed up and a fully decorrelated state is clearly reached [light (violet)]. To quantitatively extract the dynamics we must therefore combine information from all strain amplitudes. For slow dynamics, $\gamma_0 \leq 16\%$, the high-$\tau$ limit [and hence $A(q)$] can be established from the largest strain amplitude. Similarly, where decorrelation occurs rapidly ($\gamma_0 > 40\% $) the short-$\tau$ behaviour of $g^{\perp}_{\rm echo}$ [$B(q)$] can be fixed from the lowest strain, $\gamma_0 = 2.5\%$. This enables reconstruction of the echo-ISF, $f^{\perp}_{\rm echo}$, Fig.~\ref{fig:soe}(b)~(points), see ESI$^{\dag}$ Sec.~S3. From the echo-ISF the rate of microscopic rearrangements can then be extracted.

Assuming diffusive rearrangements, \textit{i.e.}~$f^{\perp}_{\rm echo}=f_D$, fitting $g^{\perp}_{\rm echo}$ using Eq.~\ref{eq:g_perp_diff} yields the diffusion coefficient of microscopic rearrangements under oscillatory flow, $D_{\rm osc}$, as a function of $\gamma_0$ [Fig.~\ref{fig:soe}(b)~(inset)]. We find the dynamics transition from being slower than the duration of the experiment ($D_{\rm osc} < \SI{D-4}{\micro\metre^2\per\second}$) at $\gamma_0 \leq 4\%$ to faster than a single oscillation cycle at $\gamma_0 \geq 100\%$ ($D_{\rm osc} > \SI{1}{\micro\metre^2\per\second}$) (solid lines). This is an increase of over 4 orders of magnitude in $D_{\rm osc}$ over $\lesssim 1.5$ decades of strain amplitude, Fig.~\ref{fig:soe}(b)[inset]~(black points). However, as the rate of rearrangement first increases the decorrelation in $f^{\perp}_{\rm echo}$ is observed not to be simple diffusion, Fig.~\ref{fig:soe}(b)~(purple). To capture the dynamics at $\gamma_0 = 6.3$ to 10\%, we use either a generalised exponential, $f^{\perp}_{\rm echo} = \exp[-(D_{\rm osc}q^2\tau)^\beta]$ (dashed lines), with a stretch exponent, $\beta <1$, or we allow the proportion of mobile particles ($n_{\rm mob}$) to be less than unity, $f^{\perp}_{\rm echo}=n_{\rm mob}\exp(-D_{\rm osc}q^2\tau)+(1-n_{\rm mob})$ (solid lines), with $1-n_{\rm mob}$ the non-ergodicity parameter. For $\beta,~n_{\rm mob} < 1$, the dynamics at these strains are heterogeneous, see ESI$^{\dag}$ Sec.~S3. A stretched exponential ($\beta<1$) implies a range of relaxation rates, with some of the dynamics on a much slower timescale. Similarly, if the ISF, $f^{\perp}_{\rm echo}$, no longer decays to 0 ($n_{\rm mob}<1$) this suggests there is a proportion of `fixed' particles that do not rearrange. In both cases, the physical interpretation is that just above the onset of rearrangements a proportion of the droplets do not rearrange over many oscillation cycles. We found that both approaches give similar quantitative measurements for measurable $D_{\rm osc}$ within experimental noise, Fig.~\ref{fig:soe}(b)~[inset].

Over the full range of strain amplitudes, the rate of rearrangements, $D_{\rm osc}$, shows a highly non-linear behaviour. A `jump' is seen between $\gamma_0 = 4\%$ ($\ll\SI{D-4}{\micro\metre^2\per\second}$) and $\gamma_0 = 6.3\%$ ($\sim \SI{D-2}{\micro\metre^2\per\second}$). As the strain amplitude increases further the increase in $D_{\rm osc}$ is approximately linear. The rate of rearrangement is captured by a critical form, $D_{\rm osc}\propto (\gamma_0-\gamma_c$), with $\gamma_c = 6\%$, Fig.~\ref{fig:soe}(b)~[inset]. This gives the rate of rearrangement as proportional to the strain accumulated above a critical microscopic yield strain, $\gamma_c$, which is the onset of irreversible motion. Below $\gamma_c$ the system deforms elastically and particles return to their original position after each cycle. Above $\gamma_c$, the particles do not return to the same position as the strain is repeatedly reversed in oscillatory flow.

Compared to the bulk rheology, the microscopic yield strain [$\gamma_c = 6\%$, Fig.~\ref{fig:rheo}(a)~(shaded region)] for the onset of irreversible deformation is lower than the bulk yield strain where the moduli cross over ($\gamma_y = 36\%$, dashed line). Instead, the onset of microscopic rearrangements is more closely linked to the end of the linear visco-elastic regime ($\gamma_{\rm LVE} = 4\%$), with the initial decrease in the storage modulus and rise in the loss modulus. As the system rapidly decorrelates, within one cycle at $\gamma_0 \geq 40\%$, we can no longer use echo-DDM to measure the microscopic dynamics of the emulsion. However, the dynamics may continue to evolve with larger deformations. To probe the yielded state under large deformations, we can instead consider steady flow and compare to the dynamics just above yielding at $\gamma_0 \gtrsim \gamma_c$.

\subsubsection{Steady flow\label{sec:flow}}

For the bulk response under a constant applied shear rate, at low shear rates, $\dot\gamma \ll \SI{1}{\per\second}$, the stress $\sigma(\dot{\gamma})$ is only weakly dependent on rate, Fig.~\ref{fig:rheo}(b)~(points). This low-shear stress is taken to represent the response to $\dot\gamma \to \SI{0}{\per\second}$, \textit{i.e.}~the solid-like response. Above a critical shear rate $\dot\gamma_c$, $\sigma$ begins to increase, transitioning to $\sigma\propto\dot\gamma^{1/2}$ scaling at $\gg \SI{10}{\per\second}$. The yield stress, $\sigma_y$, and $\dot\gamma_c$, can be extracted using a phenomenological ``three component'' model (solid line), $\sigma(\dot\gamma) = \sigma_y[1 + (\dot\gamma/\dot\gamma_c)^{1/2}] + \eta_s\dot\gamma$.\cite{caggioni2020variations} This gives $\sigma_y = \SI{31.0(4)}{\pascal}$, comparable to the oscillatory yield stress defined by moduli cross-over (dot-dashed line). In steady flow measurements, all deformation is plastic or irreversible, as shear is in a continuous direction after pre-shear (such that any yield strain is exceeded), and we henceforth refer to the irreversible shear rate, $\dot\gamma = \dot\gamma_{\rm irrev}$. In $\sigma(\dot\gamma_{\rm irrev})$, below the critical shear rate, $\dot \gamma_c = \SI{13.8(8)}{\per\second}$, while the flow is plastic the stress is dominated by the stored elastic compression of the droplets, $\sigma \approx \sigma_y$.\cite{caggioni2020variations} It is in this regime that the shear rates applied in oscillatory flow, $\dot\gamma = \omega \gamma_0$ (Sec.~\ref{sec:osc_soe}), and steady flow (see below) all lie. At $\dot\gamma_{\rm irrev} > \dot \gamma_c$, dissipation from plastic events begins to dominate as the stress increases.

\begin{figure}[tb]
    \centering
    \includegraphics{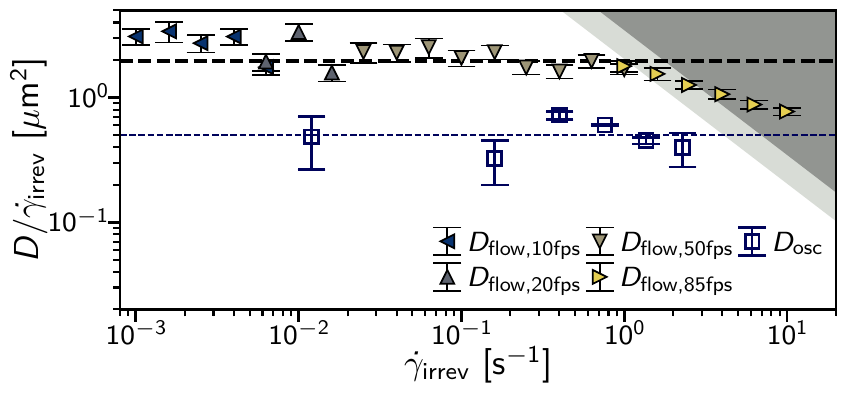}
    \caption{Comparison of steady shear and oscillatory flow. Diffusivity per unit irreversible strain \textit{vs} apparent irreversible shear rate, $D(\dot\gamma_{\rm irrev})/\dot\gamma_{\rm irrev}$. Symbols: filled, steady shear (for frame rate see inset legend); open, oscillatory flow. Shaded (grey) regions: dark, max measurable diffusivity for steady flow; light, oscillatory flow. Irreversible shear rate equal to applied shear rate, $\dot\gamma$, for steady flow and $\dot\gamma_{\rm irrev} = 4(\gamma-\gamma_c)/T$ for oscillatory flow. Critical strain for irreversible flow, $\gamma_c = 6\%$. Dotted lines: bold (black) $D_{\rm flow}/\dot\gamma_{\rm irrev} = \SI{2.0(1)}{\micro\metre^2}$; fine (blue), $D_{\rm osc}/\dot\gamma_{\rm irrev} = \SI{0.5(1)}{\micro\metre^2}$}
    \label{fig:steady}
\end{figure}

To extract the rate of microscopic dynamics in steady shear we use flow-DDM, Sec.~\ref{sec:theory_flow}, imaging at $h = \SI{10}{\micro\metre}$ over the low-rate regime ($\dot\gamma_{\rm irrev} = $~\SIrange{D-3}{10}{\per\second}). From the drift-corrected DICF in the perpendicular sector, $\bar g^{\perp}$, we extract $\bar f^{\perp}$ and from this the diffusion coefficient under steady flow, $D_{\rm flow}$, using Eq.~\ref{eq:f_diff}. The diffusivity is averaged over $q =$~\SIrange{2.0}{3.5}{\per\micro\metre}; no clear $q$ dependence is found, see ESI$^{\dag}$ Sec.~S1. When normalised by the irreversible shear rate, the rate of rearrangements per unit strain, $D_{\rm flow}/\dot\gamma_{\rm irrev}$, is almost independent of the shear rate, Fig.~\ref{fig:steady}~(a) [symbols], with $D/\dot\gamma_{\rm irrev} = \SI{2.0(1)}{\micro\metre^2}$ up to $\dot\gamma = \SI{1}{\per\second}$ (dashed line).\footnote{Fitting a power law dependence for $\dot\gamma\leq \SI{1}{\per\second}$ gives $D_{\rm flow} \propto \dot\gamma^{0.9(1)}$.} This range of $\dot\gamma_{\rm irrev}$ is below $\dot\gamma_c = \SI{14}{\per\second}$, and the flow is `slow'. A constant $D_{\rm flow}/\dot\gamma_{\rm irrev}$ is therefore consistent with simulations and experiments where the system has time to relax between each rearrangement.\cite{khabaz2020particle,vasisht2018rate} During relaxation the elastic forces between droplets are damped by viscous forces of the background solvent. Above this point the extracted strain-dependent diffusivity appears to decrease. However, these measurements are at the limit of the frame rate (dark grey shading) and we can only investigate low-rate flows, $\dot\gamma\lesssim\dot\gamma_c$, where there may not be a change in the type of rearrangement dynamics.\cite{vasisht2018rate}

To compare steady flow with oscillatory yielding we consider the relevant shear rate for oscillatory flow. In unidirectional steady shear, as stated, the applied shear is all irreversbile. However, for oscillatory flow the direction of flow continually reverses, so that elastic strain is accumulated and returned at each direction change. From $D_{\rm osc}(\gamma_0$), Fig.~\ref{fig:soe}(b)~(inset), we identified this strain as $\gamma_c$. By then assuming that all strain below this is purely elastic and all above is irreversible we can define $\dot\gamma_{\rm irrev} = 4(\gamma_0-\gamma_c)/T$, where we divide the total irreversible strain accumulated in an oscillation cycle by the period.\footnote{Over a cycle, a strain of $2\gamma_0$ occurs in both directions. On reversal of the flow direction there is a recovery of $\gamma_c$ elastic strain from the previous flow direction and accumulation of $\gamma_c$ elastic strain in the new direction; this gives $\gamma_{\rm irrev} = 2(\gamma_0-\gamma_c)$ in each direction and hence $\dot\gamma_{\rm irrev} = 4(\gamma_0 - \gamma_c)$ for $\gamma_0 > \gamma_c$ or $\dot\gamma_{\rm irrev} = 0$ for $\gamma_0 < \gamma_c$.} The rate of rearrangement with irreversible strain in oscillatory shear then appears independent of rate, Fig.~\ref{fig:steady}~(large open symbols), although with only a limited number of points. This is consistent with the system yielding to a steadily flowing state ($D_{\rm osc} \propto \dot\gamma_{\rm irrev}$), rather than undergoing an evolution over larger strains.\cite{rogers2012sequence} This gives a qualitative view of the microscopic dynamics of yielding in oscillatory flow. We note that $D_{\rm osc}\approx D_{\rm flow}/4$ potentially suggests that shear-induced rearrangements in oscillatory shear are quantitatively different to those in continuous shear. However, these microscopic shear-induced diffusivities are written in terms of the macroscopically applied strain, which we have implicitly assumed to be uniform across the height, $h$, of the sample. We show in the next section that this approximation is not fulfilled.

\subsubsection{Flow velocity profile}

\begin{figure}[tb]
    \centering
    \includegraphics[width=\columnwidth]{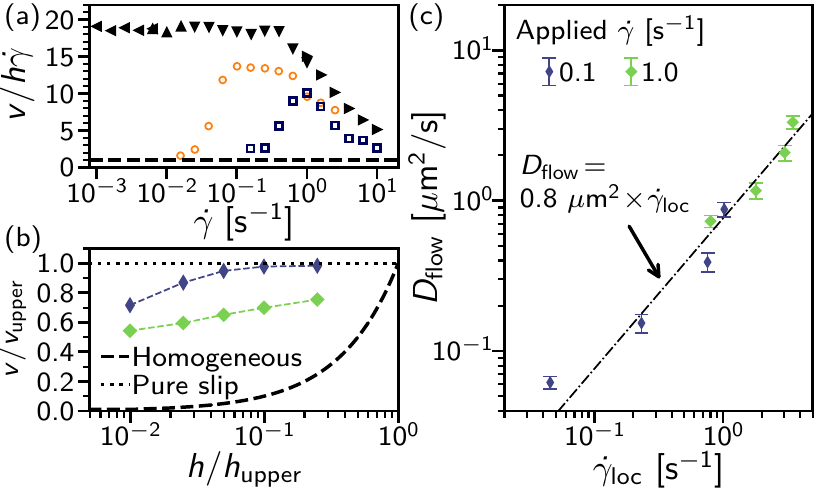}
    \caption{Flow profile. (a)~Comparison of steady shear (filled symbols) and oscillatory flow [open: \SI{0.1}{\hertz}, light (orange); \SI{1}{\hertz} dark (blue)] at $h = \SI{10}{\micro\metre}$ with changing applied shear rate. Velocity normalised by velocity for homogeneous flow (dashed line), $v/h\dot\gamma$ (for oscillatory flow $\dot\gamma = 2\pi \gamma_0/T$). (b)~Velocity for steady flow with changing sample depth, $h$, normalisation by velocity of upper cone at imaging radius ($v_{\rm upper}$) and height ($h_{\rm upper}$). For applied $\dot\gamma$ see legend in (c). Lines: dashed, homogeneous flow; dotted, pure slip. (c)~Measured diffusivity, $D_{\rm flow}$, with local shear rate, $\dot\gamma_{\rm loc}$, derived from gradient in $v(h)$. Dot-dashed line, $D_{\rm flow}= 0.8(1)\si{\micro\metre^2} \times\dot\gamma_{\rm loc}$; $h = \SI{50}{\micro\metre}$ not shown due to insufficient signal strength}
    \label{fig:height}
\end{figure}

In complex fluids even when the stress is applied uniformly, the system may flow inhomogeneously. This may be a property of the sample, as in shear banding where the fluid can exist at, \textit{e.g.}~different shear rates under the same stress.\cite{olmsted2008perspectives} Alternatively, in pure slip a layer of the background fluid may lubricate flow between the droplets and the shearing surface.\cite{meeker2004slip} As part of the analysis process, flow-DDM also gives information on the mesoscale dynamics, \textit{i.e.}~the local flow velocity as a function of the height of the focal plane within a sample, $v(h)$, via the method of phase dynamic microscopy.\cite{colin2014fast} Up to this point we have discussed deformation in terms of the average strain over the whole gap. However, for $\dot\gamma \lesssim \SI{1.0}{\per\second}$, the extracted velocity from flow-DDM, $v$, is $\approx 20\times$ the expected velocity for a uniform shear rate across the gap, $h\dot\gamma$, Fig.~\ref{fig:height}(a). This suggests that as the system yields the flow is localised near the lower plate. At higher $\dot\gamma$, the normalised flow speed decreases and approaches one, suggesting that the flow becomes less localised and the critical shear rate, $\dot\gamma_c$, may plausibly be related to the onset of the formation of a homogeneously flowing state, reminiscent of banding in thixotropic systems.\cite{ovarlez2009phenomenology}  

As we use optically sectioned confocal microscopy, we can investigate the flow profile in steady shear by imaging at multiple heights: $2,~5,~10,~20$ and \SI{50}{\micro\metre} for two applied shear rates $\dot\gamma = 0.1$ and \SI{1.0}{\per\second}, Fig.~\ref{fig:height}(b). The extracted velocity, $v(h)$, normalised by the velocity of the upper surface of the geometry $v_{\rm upper}$ at $h_{\rm upper} = \SI{200}{\micro\metre}$, shows the velocity profile. At $\dot\gamma = \SI{0.1}{\per\second}$ [thin (purple) diamonds], shear is highly localised near the lower surface. The lowest height imaged, $h = \SI{2}{\micro\metre}$, is comparable to the surface roughness. The speed of the upper geometry is reached at $h = \SI{20}{\micro\metre}$, above this the sample is rotating as a solid. Between these two points ($\approx 10\%$ of the gap) the sample is sheared and the local shear rate, $\dot\gamma_{\rm loc}$ can be estimated from the gradient in $v(h)$. For $h\leq\SI{10}{\micro\metre}$, $\dot\gamma_{\rm loc} > \dot\gamma$. With a higher applied shear rate, $\dot\gamma = \SI{1}{\per\second}$, Fig.~\ref{fig:height}(b)~[fat (green) diamonds], the sheared region extends further into the bulk of the sample, with $v(h \! = \! \SI{50}{\micro\metre})<v_{\rm cone}$. Although shear is localised compared to homogeneous flow, \textit{cf.}~symbols and dashed line [Fig.~\ref{fig:height}(b)], in contrast to pure slip (dotted line) droplets are still sheared past one another within a finite layer.

Within the sheared region, we found the extracted diffusivity under steady flow is proportional to the local shear rate, $D_{\rm flow} \approx \dot\gamma_{\rm loc}\times \SI{0.8}{\micro\metre^2}$, Fig.~\ref{fig:height}(c). To compare this to the diffusivity at $h = \SI{10}{\micro\metre}$ as a function of applied rate, Fig.~\ref{fig:steady}~(filled symbols), we can estimate $\dot\gamma_{\rm loc}(h=\SI{10}{\micro\metre}) \approx 2\dot\gamma$. Neglecting the weak dependence on applied shear rate, the diffusivity per unit strain [Fig.~\ref{fig:steady}~(filled symbols] can be estimated in terms of the local shear rate: $D_{\rm flow} \approx \SI{1.0}{\micro\metre^2} \times \dot\gamma_{\rm loc}$, far closer to the diffusivity extracted with changing $h$, Fig.~\ref{fig:height}(c). Therefore, for steady flow at low $\dot\gamma$ there is a single well-defined microscopic diffusivity per unit local strain. However, this is accompanied by complex and changing shear-localisation dynamics.

\begin{figure}
    \centering
    \includegraphics{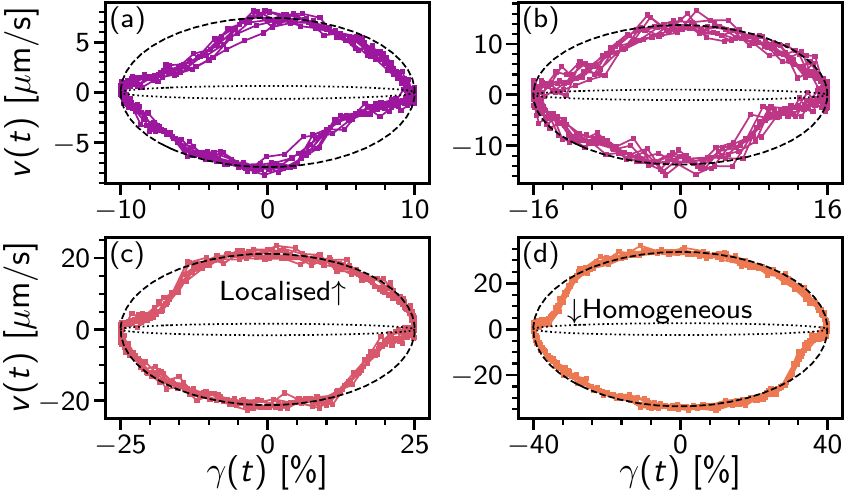}
    \caption{Local velocity in oscillatory shear. (a)~Parametric plot of extracted velocity, $v(t)$ vs applied strain, $\gamma(t)$ at \SI{0.1}{\hertz} over 8 cycles for $\gamma_0 = 10\%$. Time increases in a clockwise direction. Lines: dotted, homogeneous flow velocity, $v = \dot\gamma h$; dashed, localised, $v \gg \dot\gamma h$ based on value in Fig.~\ref{fig:height}(a); solid, linking $v(t)$ data. (b)~$\gamma_0=16\%$. (c)~$\gamma_0 = 25\%$. (d)~$\gamma_0=40\%$}
    \label{fig:osc_velocity}
\end{figure}

With this insight we return to oscillatory flow. Using the velocity measurement technique of flow-DDM on short subsections of several oscillatory cycles, we estimate $v(t)$ through out a cycle, with the maximum velocity shown in Fig.~\ref{fig:height}(a). For \SI{1.0}{\hertz}~[(dark (blue) squares], at the lowest strains, $\gamma_0\leq 4\%$ (and correspondingly $\dot\gamma = \omega \gamma_0 \leq \SI{0.25}{\per\second}$)\footnote{As both elastic and plastic strain causes bulk motion, we return to the conventionally defined oscillatory shear rate.}, the velocity is consistent with a near homogeneous strain across the gap ($\dot\gamma \approx \dot\gamma_{\rm loc}$). As $\gamma_0$ increases $\geq 6.3\%$ ($\dot\gamma \geq \SI{0.4}{\per\second}$), $v/h\dot\gamma$ increases to $\gg 1$ and the applied strain is localised near the lower plate. However, the degree of localisation is consistently lower than for steady flow (\textit{cf.}~open and filled symbols), even at $\gamma_0 \geq 100\%$. A comparable sequence can be seen from velocimetry on a limited number of cycles at a lower frequency, \SI{0.1}{\hertz}[(orange) circles].

At \SI{0.1}{\hertz} we can also observe the strain-dependent shear localisation \emph{within} a cycle, Fig.~\ref{fig:osc_velocity}. To examine intracycle dynamics, we parametrically plot the applied strain, $\gamma(t)$, and the locally measured velocity, $v(t)$ (shaded symbols), for a range of strains (10\% $\leq \gamma_0 \leq 40\%$) above $\gamma_c$. For most of the cycle the system is strongly shear-localised ($v \gg h\dot\gamma$, black dashed lines). However, immediately upon reversal (\textit{i.e.}~about $v(t)=\SI{0}{\micro\metre\per\second}$) the velocity is lower for $\approx 6\% (=\gamma_c)$ before rising to the banded state in another $\sim 5\%$. The transient intracycle velocity is then consistent with the intercycle measurements of peak velocity increasing with $\gamma_0$, Fig.~\ref{fig:height}(a)~(open symbols): as strain is applied, the system first deforms uniformly ($\lesssim \gamma_c$) before localising.

\section{Discussion and Conclusions\label{sec:conc}}

Using a combination of echo-DDM and flow-DDM we have demonstrated how differential dynamic microscopy can reveal the microscopic shear-induced dynamics in flowing complex fluids. As an example, we have investigated rearrangements of droplets in a concentrated silicone oil emulsion to probe the yielding transition on a microscopic level. Insight is gained into the complete yielding process by a comparison of echo-DDM for oscillatory shear and flow-DDM analysis of steady shear. Together, these reveal both the diffusive dynamics and the local flow velocity, which ultimately enables us to relate the changing local processes to the bulk rheology.

From echo-DDM analysis on oscillatory shear, we measured a critical shear strain for the onset of rearrangements, $\gamma_c\approx 6\%$. At small strains, $\gamma_0 < \gamma_c$ the system deforms elastically. On the microscopic level in echo-DDM, where the system is probed each time it returns to the same bulk position, it is seen that droplets do not rearrange even over 100 cycles, Fig.~\ref{fig:soe}(a)~[dark line]. The shear-induced diffusivity $D_{\rm osc}$ is below the minimum measurable rate set by the length of the experiment, Fig.~\ref{fig:soe}(b)~[inset]. This elastic deformation is close to homogeneous, Fig.~\ref{fig:height}(a)~(open symbols), where the strain is near uniform across the gap ($v \approx \dot\gamma h$). In the bulk rheology this corresponds to the linear viscoelastic regime, Fig.~\ref{fig:rheo}, with a constant elastic modulus ($G^{\prime}$) that is much larger than the loss modulus ($G^{\prime\prime}$): the emulsion behaves as a solid.

As the strain amplitude increases droplets must rearrange. For $6.3\% \leq \gamma_0 \leq 10\%$, the system decorrelates over multiple cycles, Fig.~\ref{fig:soe}(b). Just above $\gamma_c$, the dynamics are highly heterogeneous. Fitting to a stretched exponential or including a proportion of non-mobile particles ($1-n_{\rm mob}$) that limits full decorrelation of the ISF, shows that there is a population of droplets where the dynamics remain slow. This aligns with previous observations of a super-mobile population under oscillatory shear measured stroboscopically,\cite{knowlton2014microscopic} although heterogeneity could also be plausibly caused by the droplet polydispersity, Fig.~\ref{fig:confocal}(c). As the strain amplitude is increased up to $\gamma_0 =40\%$, the rate of rearrangements increases, until the droplets have rearranged within one cycle, Fig.~\ref{fig:soe}(b)~[inset]. The transition in the behaviour at $\gamma_0= 6.3\%$ is accompanied by a non-linear `jump' in the measured diffusivity, or rate of rearrangements. The dependence of $D_{\rm osc}$ is instead described by proportionality to the plastic strain, $\gamma_0 - \gamma_c$, which is that above the strain that can be accommodated elastically. So, at small strain, deformation is rearrangement free (on experimental timescales) and elastic until the critical strain is reached, after which all further strain is plastic and drives rearrangement events. In general this separation may not be as clear, highlighting the importance of measurements at the microscopic level.\cite{kamani2021unification}

The onset of irreversible, plastic deformation is accompanied by shear localisation from $\gamma_0 = 6.3\%$ to 16\%, this is apparent from the flow velocity, Fig.~\ref{fig:height}(a)~(open symbols). During a low-frequency cycle this transition from homogeneous elastic flow to localised plastic flow can be seen within a single cycle upon direction reversal at large amplitude ($\gamma_0 \gg \gamma_c$), Fig.~\ref{fig:osc_velocity}. With a region of high shear rate near the lower plate the local shear rate at $h$, $\dot\gamma_{\rm loc}$, is no longer given by just the macroscopic strain. The yielding process that begins with departing the linear elastic regime at $\gamma_{\rm LVE}$, Fig.~\ref{fig:rheo}(a), then corresponds to shear localisation dynamics. Bulk yielding defined by moduli crossing over, $\gamma_{y} = 36\%$, is therefore after a localised flow profile is established within a cycle.

Under steady flow, a velocity profile can be measured by recording at varying heights and using the velocity extracted from flow-DDM, Fig.~\ref{fig:height}(b). The local shear rate is given by the velocity gradient in $h$. The diffusivity extracted from flow DDM for $\SI{2}{\micro\metre} \leq h \leq \SI{20}{\micro\metre}$ is proportional to the local shear rate, $D_{\rm flow} = \dot\gamma_{\rm loc} \times \SI{0.8}{\micro\metre^2}$. This behaviour is comparable to simulations of soft particle systems where droplets have time to relax between each rearrangement event at slow shear rates, $\dot\gamma < \dot\gamma_c$.\cite{khabaz2020particle} At a given lengthscale, this gives an irreversible strain for the system to rearrange, $\gamma_R = \dot\gamma_{\rm loc}\tau_D$ with $\tau_D=1/q^2D_{\rm flow}$ the characteristic time of diffusion in steady flow. Using $q = \SI{2}{\per\micro\metre}$ (lower $q$ limit for echo-DDM and flow-DDM), $\gamma_R = (q^2\times\SI{0.8}{\micro\metre^2})^{-1} = 30\%$, for a lengthscale ($L = 2\pi/q \approx \SI{3}{\micro\metre}$) corresponding to the largest droplet size, Fig.~\ref{fig:confocal}(c). At higher shear rates, above our measurement range, $\gamma_R$ is relevant to the onset of changes in $\sigma(\dot\gamma)$.\cite{khabaz2020particle} 

In oscillatory flow the peak velocity at $h=\SI{10}{\micro\metre}$ is below that of steady shear at the same applied rate, \textit{cf.}~open and filled symbols, Fig.~\ref{fig:height}(a). Therefore, the shear localisation dynamics are different in oscillatory and steady flow. We cannot then relate the macroscopic plastic strain, $\gamma_0-\gamma_c$ to a local shear rate in the diffusivity relation for echo-DDM at \SI{1}{\hertz}, $D_{\rm osc} = \SI{2}{\micro\metre^2\per\second} \times (\gamma_0-\gamma_c)$ Fig.~\ref{fig:soe}(b)~[inset (line)]. If the relation from steady shear holds, $D_{\rm flow} = \SI{0.8}{\micro\metre^2}\times \dot\gamma_{\rm loc}$, the local shear rate at $h=\SI{10}{\micro\metre}$ in oscillatory shear is less than the macroscopic irreversible strain, $\gamma_{\rm loc} = 0.6\dot\gamma_{\rm irrev}$. Further work probing oscillatory flow at multiple heights would test this prediction. 

Typically, shear localisation, in the form of banding, is reported for attractive emulsions where an excess of the surfactant (SDS) causes depletion interactions.\cite{becu2005concentrated,becu2006yielding} Diluting the emulsion in the background solvent used shows separate droplets rather than clusters, see ESI$^{\dag}$ Sec.~S4. Attraction between droplets is therefore minimal. Banding in systems with only soft elastic interactions has become an area of recent exploration.\cite{vasisht2020computational,ozawa2018random} These suggest how the system is prepared or annealed is crucial as it can control a transition from ductile to brittle yielding with increased thermal annealing. However, our experimental system is prepared by mechanical pre-shearing. Understanding the impact of pre-sheared states on the banding dynamics in athermal amorphous materials is an active area of theoretical work\cite{das2020unified} and thus their future experimental characterisation using a combination of echo-DDM and flow-DDM appears of great interest.

In conclusion, we have shown how DDM can be adapted to study rearrangements in flowing complex fluids. For oscillatory shear, we have developed echo-DDM and validated it using a sample whose dynamics do not depend on the applied shear. Carefully minimising and accounting for the impact of flow was found to be critical. Using a combination of echo-DDM and flow-DDM on oscillatory shear and steady shear, respectively, we show it allows a comprehensive microscopic picture of yielding in non-Newtonian complex fluids. We demonstrate it using a concentrated silicone-oil emulsion imaged with confocal microscopy. Using the extracted diffusivities ($D_{\rm osc}$ and $D_{\rm flow}$) quantifying the shear-induced microscopic rearrangement under oscillatory and steady flow, and the measured flow velocities, yielding was revealed as a two-step process. In the first step the particles begin to arrange, with heterogeneous behaviour close to this threshold. In the second step the applied shear becomes localised near the lower plate, leading to macroscopic yielding based on the bulk rheology. As both echo-DDM and flow-DDM are based on differential dynamic microscopy, there are several advantages to be exploited. Firstly, it is independent of microscopy method: the local flow velocity can be extracted without confocal sectioning\cite{richards2021particle} and diffusivity has been measured with multiple methods.\cite{bayles2016dark,cerbino2008differential,lu2012characterizing} Secondly, particle resolution is not needed and with this comes a large field of view that allows rapid data acquisition suitable for a high-throughput technique\cite{martinez2012differential} and information on multiple lengthscales.\cite{cerbino2017perspective} This suite of analysis methods and experimental protocols can therefore illuminate the yielding process in materials that are not amenable to particle-level methods, such as suspensions with complex interactions.\cite{guy2018constraint,richards2021turning}

\section*{Conflicts of Interest}
There are no conflicts to declare.

\section*{Acknowledgements}
This project has received funding from the European Research Council (ERC) under the European Union's Horizon 2020 research and innovation programme (grant agreement \ftextnumero{s}~731019 and 862559), European Soft Matter Infrastructure (EUSMI) and Novel Characterisation Platform for Formulation Industry (NoChaPFI). All data used are available via Edinburgh DataShare at \url{https://doi.org/10.7488/ds/XXXX}.

\balance
\providecommand*{\mcitethebibliography}{\thebibliography}
\csname @ifundefined\endcsname{endmcitethebibliography}
{\let\endmcitethebibliography\endthebibliography}{}

\end{document}